# Detection of a Dipole in the Handedness of Spiral Galaxies with Redshifts $z \sim 0.04$

Michael J. Longo[1]

Department of Physics, University of Michigan, Ann Arbor, MI 48109, USA

**Abstract**
   A preference for spiral galaxies in one sector of the sky to be left-handed or right-handed spirals would indicate a parity violating asymmetry in the overall universe and a preferred axis. This study uses 15158 spiral galaxies with redshifts <0.085 from the Sloan Digital Sky Survey. An unbinned analysis for a dipole component that made no prior assumptions for the dipole axis gives a dipole asymmetry of $-0.0408 \pm 0.011$ with a probability of occurring by chance of 7.9 x $10^{-4}$. A similar asymmetry is seen in the Southern Galaxy spin catalog of Iye and Sugai.  The axis of the dipole asymmetry lies at approx. $(l, b) = (52°, 68.5°)$, roughly along that of our Galaxy and close to alignments observed in the WMAP cosmic microwave background distributions. The observed spin correlation extends out to separations ~210 Mpc/h, while spirals with separations < 20 Mpc/h have smaller spin correlations.

## 1. Introduction

A basic assumption of essentially all cosmological models and general relativity is the "Cosmological Principle" that over large enough distance scales the Universe is homogeneous and isotropic. This paper presents strong evidence for a parity-violating special axis as demonstrated by a dipole in the distribution of spiral galaxy handedness for redshifts <0.085.

On the smallest scales, a parity-violating asymmetry was found in the angular distribution of electrons in the beta decay of spin-oriented $^{60}$Co, confirming the proposal by Lee and Yang [1] that parity is violated in weak decays.  On the molecular scale, there is a large predominance of left-handed amino acids over right-handed ones in organisms, the origin of which is still not well understood.  It is reasonable to ask if nature exhibits such an asymmetry on the largest scales.

There is already significant evidence for a cosmic parity violation.  Kim and Naselsky [2] show that the odd-parity preference of the cosmic microwave background (CMB) power spectrum from the Wilkinson Microwave Anisotropy Probe (WMAP) is anomalous at the 4-in-1000 level. A. Gruppuso et al. [3] have also analyzed the WMAP7 temperature and polarization maps and find a parity anomaly at the 99.5% confidence level. S. Alexander [4] argues that a parity violating extension to general relativity can explain the WMAP anomalies and provide a mechanism for inflationary leptogenesis.

---
[1] email: mlongo@umich.edu

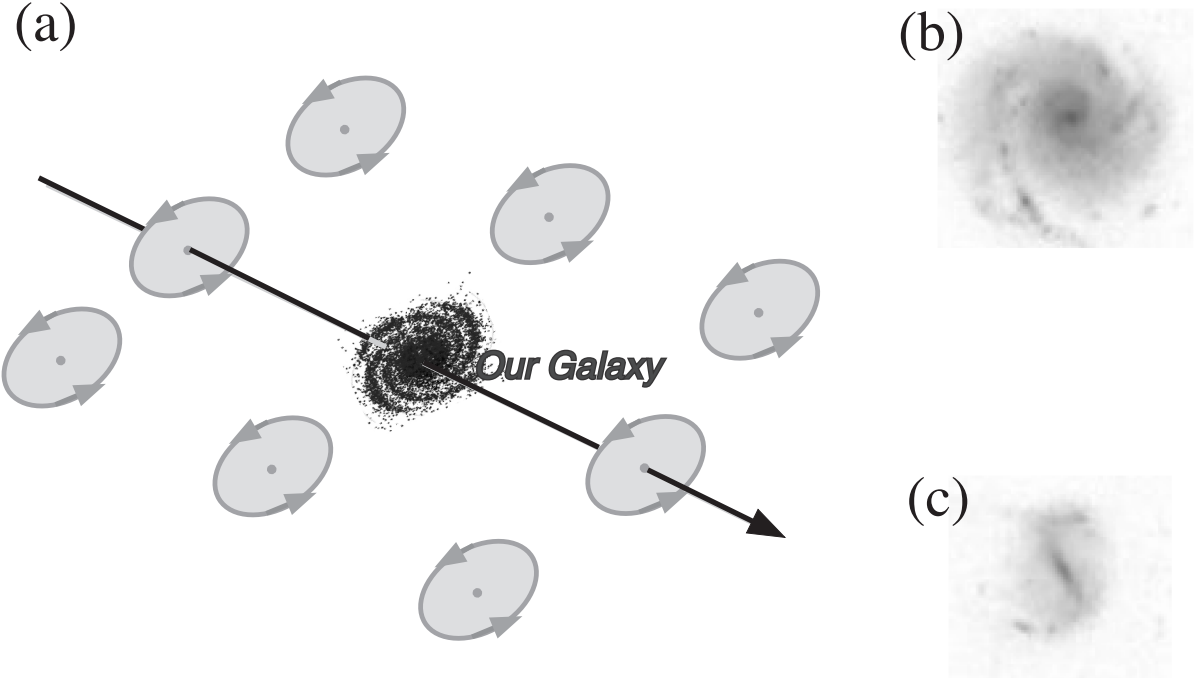

FIG. 1(a)–A hypothetical universe with all galaxies having the same handedness. Note that galaxies in one hemisphere would appear to us to be right-handed and in the opposite hemisphere left-handed. (b) A "typical" spiral galaxy from the SDSS. This one is defined as having right-handed "spin". (c) A left-handed two-armed spiral galaxy.

Spiral galaxies with a well-defined handedness offer a means of testing the possibility of a cosmic parity violation. Ideally the signal for such an asymmetry would be an excess of one handedness or "spin" in a large region of the sky and a similar excess of the other handedness in the opposite direction (i.e., a dipole), as shown schematically in Figure 1. The preponderance of data in the northern Galactic hemisphere, as well as the masking of much of the sky by dust in the Milky Way, complicates the search for such an effect. However, the spiral handedness technique has important advantages over other cosmic anisotropy tests in that it is not biased by the incompleteness of the maps or by atmospheric or instrumental effects, which cannot turn right-handed spirals into left-handed. One has to be careful of an overall bias due to a preference toward assigning left-handed or right-handed. Such a bias would show up as a "monopole". Precautions against such a left/right bias will be discussed below.

In an earlier study [5], I used galaxies from the SDSS Data Release 5 data base [6] that contains ~40,000 galaxies with spectra for redshifts $z < 0.04$. In this study I use the DR6 data base [7] with ~230,000 galaxies with $z < 0.085$. A few percent of these are spiral galaxies with identifiable handedness that can be used in the study.

## 2. The Analysis

Objects classified as "galaxies" in the SDSS DR6 database were used in this analysis. A list of galaxies that had spectroscopic redshifts less than 0.10 was obtained from the SDSS DR6 web sites, cas.sdss.org/dr6 and casjobs.sdss.org. Spiral galaxies are typically bluer than elliptical ones. Strateva et al. [8] show that elliptical galaxies can be separated from spirals fairly cleanly by a $z$-dependent cut on ($U$-$R$) where $U$ and $R$ are the apparent magnitudes for the ultraviolet (354 nm) and red (628 nm) bands respectively (York et al. [9]). A conservative cut to enhance the fraction of spirals was therefore made by requiring that ($U$-$R$) < 2.85. Galaxy images from the resulting list of ~200,000 galaxies were then looked at by a team of 5 scanners.

Individual RGB images of the galaxies from the list were acquired from the SDSS web site and displayed to the scanners using an HTML/JAVA program. Scanners were assigned small $z$ slices at random, and the scanning was done in random order with respect to right ascension, $\alpha$, and declination $\delta$ so that any scanning bias could not cause a systematic bias in the ($\alpha$, $\delta$) distributions of the handedness, only a possible overall bias in the complete sample. The scanners had only 3 choices: <u>L</u>eft, <u>R</u>ight, or <u>U</u>nclear, where Left ≡ ↺ and Right ≡ ↻. No attempt to otherwise classify the galaxies was made. Scanners were instructed to classify galaxies as Unclear unless the handedness was clear. Overall, about 15% of the galaxies were classified as having recognizable handedness ($L$ or $R$). No further analysis of the $U$'s was done.

The HTML program mirrored half of the images at random to avoid scanning biases favoring a particular handedness. The scanners had no visual cue as to whether the image was mirrored. This essentially eliminated systematic effects due to scanner biases since hard-to-classify galaxies had an equal chance of being mislabeled as $L$ or $R$, and any inherent scanner bias would be equally likely to turn $L$ into $R$ or vice versa. In general, incorrectly classified spirals only dilute any real left-right asymmetry with no overall bias.

To reproduce the earlier study [5] as closely as possible, I required that the green magnitude be <17 for redshifts $z < 0.04$. Beyond $z=0.04$ the magnitude limit was increased to 17.4. The handedness of galaxies fainter than magnitude 17.4 and galaxies with $z > 0.085$ were generally difficult to classify, so these were not used. As in the previous studies, a cut was made to remove the bluest galaxies that tend to be those with recent star formation initiated by a collision. This required ($U$-$Z$) > 1.6, where $Z$ is the apparent magnitude in the far infrared; it removed 1.9% of the *L+R* sample. A cut to remove the reddest galaxies, (*U-Z*) < 3.5 was also made; this removed 2.7% of the sample, leaving 15158.

Most of the SDSS DR6 data are in the northern Galactic cap centered at $\alpha$~192°, $\delta$~ 27° in equatorial coordinates. In right ascension the coverage is roughly 120°< $\alpha$< 240° and in declination -5°< $\delta$ <63°. In the southern Galactic hemisphere there is only coverage in 3 narrow bands in $\delta$ near $\delta$=−10°, 0°, and 14°, each about 4° wide, with $\alpha$ between roughly –60° and 60°.

None of these cuts or the incomplete coverage of the survey would be expected to cause a bias between left- and right-handed spirals.

## 3. Results

A plot of asymmetries $\langle A \rangle \equiv (R - L)/(R + L)$, binned in 30° sectors of right ascension and 0.01 slices in *z* for *z*<0.085, is shown in Fig. 2. Positive $\langle A \rangle$ are shown in red and negative ones in blue. The larger numbers near the perimeter give the net asymmetry for the entire right ascension sector. The black numbers in parentheses next to them give the total number of galaxies in that sector. The $\sigma$ are determined from standard normal distribution statistics, $\sigma(N) = \sqrt{N}$, which gives $\sigma(\langle A \rangle) = 1/\sqrt{R+L}$. There is an apparent excess of left-handed spirals in the sectors for 150°< $\alpha$<240° and a complementary excess of right-handed in the opposite hemisphere, though there are only 1/7$^{th}$ as many galaxies there.

The incompleteness of the survey, especially in $\delta$, makes a complete multipole analysis of the asymmetry data of dubious value. In any case a preferred spiral handedness implies a dipole component and the lack of a monopole (bias), so I restrict my analysis to these two terms.

*Bias*–The galaxies were scanned in random order with respect to ($\alpha$, $\delta$, *z*), so no ($\alpha$, $\delta$) dependent bias is possible. In addition, half the galaxies were randomly mirrored during scanning with no visual cues as to the mirroring, and precautions against left-right bias were taken in the web interface used by the scanners. The best check for an overall bias is to look at the complete scanned sample that included galaxies at larger *z* and fainter luminosities as well as some that were scanned more than once. This sample included 25612 galaxies and gave *R*=12707, *L*= 12905 and an overall asymmetry of −0.0077 ± 0.0062, even though this sample included almost twice as many galaxies in the 150°< $\alpha$<240° sector with its apparent excess of *L* spirals as in all other sectors combined (Fig. 2). If the 150°<$\alpha$<240° sector is removed, the asymmetry becomes +0.0133±0.010, consistent with no bias or a small positive one.

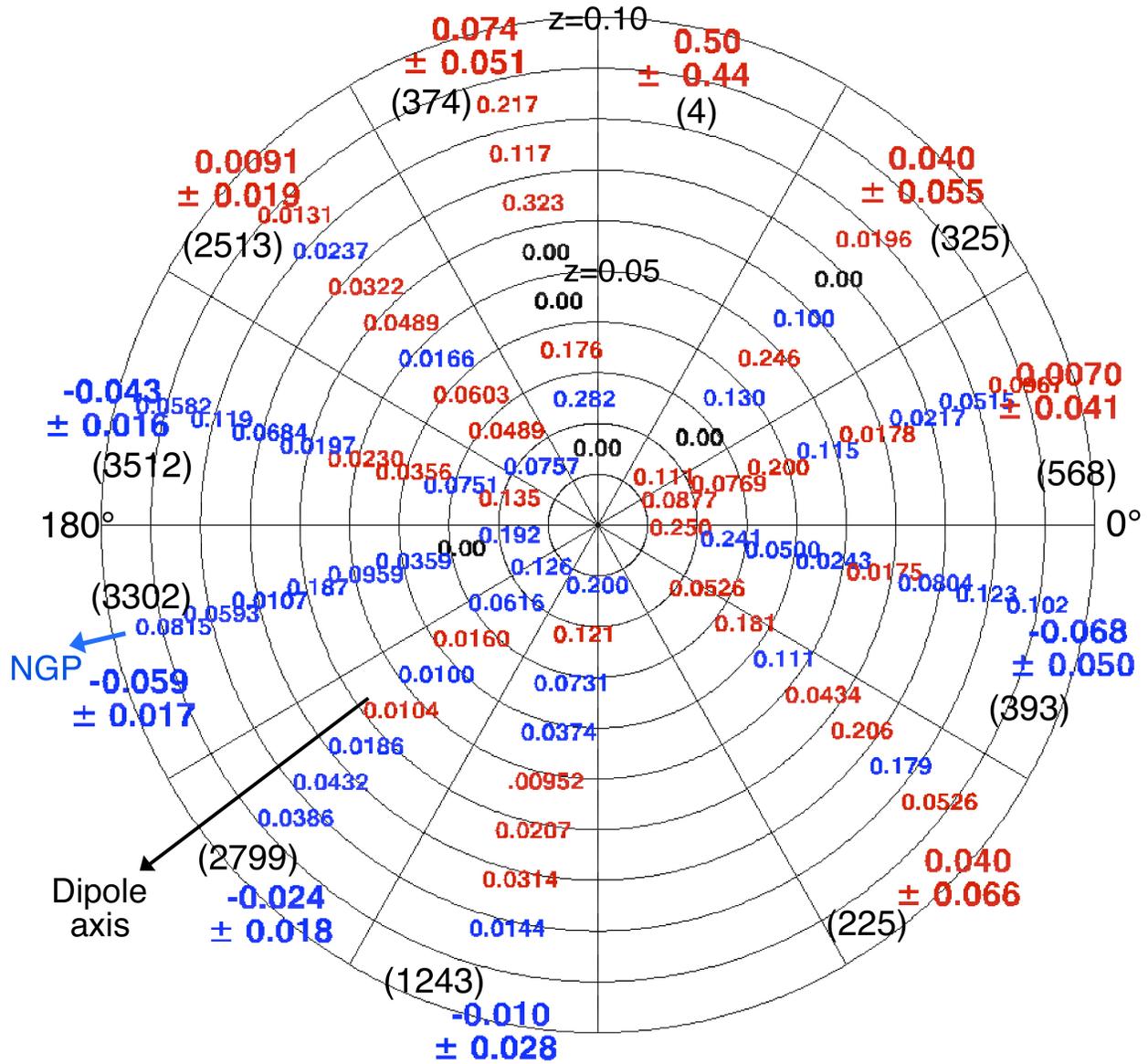

FIG. 2 – Polar plot of net asymmetries $\langle A \rangle$ in 30° sectors in right ascension and slices in $z$. Segments with positive $\langle A \rangle$ are indicated in red and negative $\langle A \rangle$ in blue. The $\langle A \rangle$ for segments with <10 galaxies are not shown. The larger numbers near the periphery give the overall asymmetry for that sector; the black numbers in parentheses are the total number of spiral galaxies in the sector. The NGP is the north pole of our Galaxy, so that the left half of the plot corresponds roughly to the northern Galactic hemisphere. The black arrow shows the most probable dipole axis. Declinations between −19° and +60° were used.

*The Dipole*–A real large-scale spiral asymmetry would exhibit itself as a dipole with a $\cos\gamma$ dependence where $\gamma$ is the space angle between the position of the galaxy and the axis of the dipole. This $\cos\gamma$ behavior is not affected by the incompleteness of the sample, but the sparseness of the data in one hemisphere makes fitting the $\cos\gamma$ dependence more difficult. To investigate a dipole, the complete sample of 15158 galaxies without position cuts was used. No a priori assumptions about the direction of the dipole axis or its magnitude were made, and no binning of the data was used in the analysis, so that the results do not depend on the choice of bins. First a possible axis was chosen from within the SDSS coverage, roughly $120° < \alpha < 240°$ and $-5° < \delta < 63°$. The $\gamma_i$ of each of the galaxies were calculated for that axis and the observed handedness of +1 or -1 was assigned to each galaxy. The 15158 points were then fitted to an $a\cos\gamma_i$ dependence. The $(\alpha_A, \delta_A)$ of the axis was varied stepwise to find the axis that gave the minimum $\chi^2/dof$. The best fit was found at $(\alpha_A, \delta_A) = (217°, 32°)$, or $(l, b) = (52°, 68.5°)$ in Galactic coordinates.

Fig. 3 shows the variation of $(\chi^2 - dof)$ with $\alpha_A$ for $\delta_A = 32°$, the axis declination that gave the minimum $\chi^2$. The x's are the values obtained for axes that are in the plane orthogonal to the best-fit axis; these show a much smaller variation in $\chi^2$ and their average is close to 0. Note that while $\chi^2$ is a very well behaved function of $\alpha_A$ and $\delta_A$, its distribution is very unlike that of the usual $\chi^2$ distribution because the handedness can only take discrete values of +1 or –1 while the usual $\chi^2$ distribution is expected only for data that have a normal distribution. Thus in this case the statistical significance of the observed $\chi^2$ had to be determined numerically for the actual galaxy sample. The $\chi^2$ distribution that would be expected if the handednesses were distributed randomly was determined by generating $4 \times 10^5$ samples of the 15158 galaxies with the handedness of each galaxy randomly assigned to either +1 or –1 with equal probability. In order to determine the probability that the observed dipole term could have occurred by chance for an arbitrarily chosen axis within the SDSS survey, the $\chi^2$ for each of 100 axes randomly chosen from among the $(\alpha_A, \delta_A)$ of the 15158 galaxies was calculated and the axis that gave the minimum $\chi^2$ was found. The distribution of $\chi^2_{min}$ was then studied for the $4 \times 10^5$ randomized handedness samples. This procedure of choosing axes randomly from among the 15158 spirals guaranteed that the simulated samples covered the same angular range as that searched for the real sample. The $(dof - \chi^2_{min})$ distribution for the $4 \times 10^5$ handedness samples is shown in Fig. 4. The arrow shows the value of $(dof - \chi^2) = 13.356$ for the observed handedness assignments and $(\alpha_A, \delta_A) = (217°, 32°)$; the probability of finding that value or greater by chance is $7.9 \times 10^{-4}$. Many checks were made to verify this probability estimate. Samples with 20, 50, or 200 possible axes, rather than 100, gave substantially the same probability distributions as in Fig. 4. For 10 tries with the actual spins, the lowest $(dof - \chi^2)$ from among 100 possible axes was always >13.2.

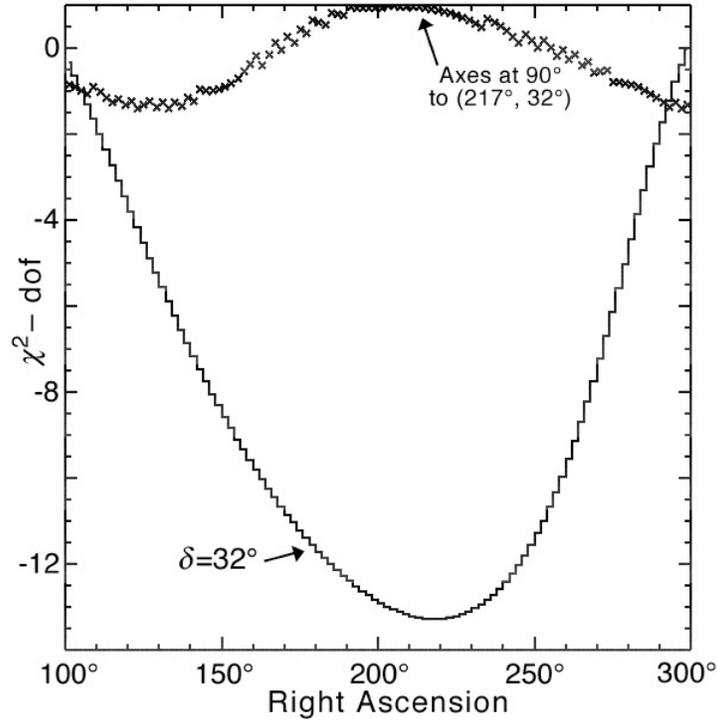

FIG. 3 – Variation of $\chi^2-dof$ with $\alpha_A$ for $\delta_A=32°$. The x's are for axes at 90° to the best-fit axis.

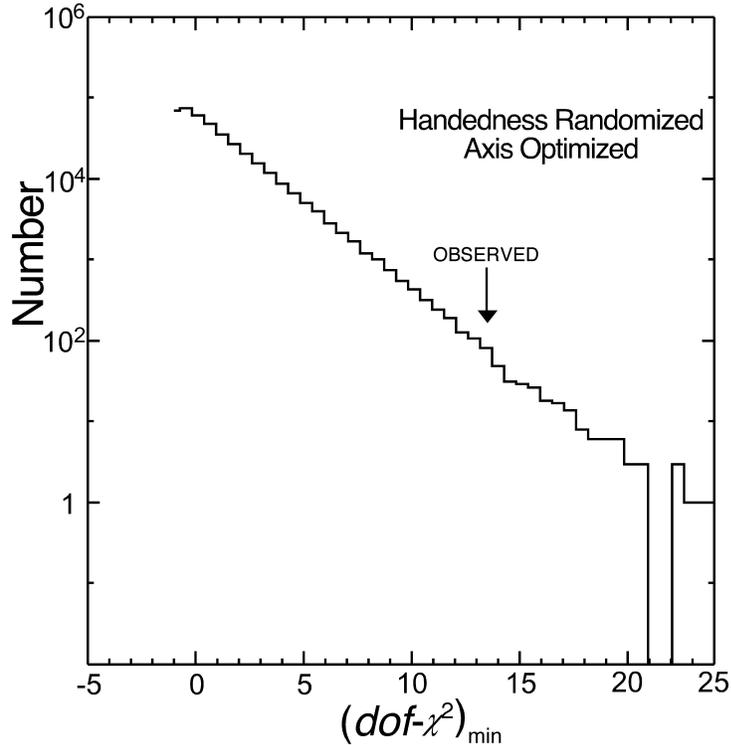

FIG. 4 – Probability of obtaining a particular value of $(dof-\chi^2_{min})$ for $4 \times 10^5$ samples of the 15158 galaxies with <u>randomized</u> handednesses. The lowest $\chi^2$ from 100 axes randomly chosen within the SDSS survey was used. The arrow shows the value of $\chi^2=13.356$ for the actual handedness assignments for the best-fit axis at $(\alpha_A, \delta_A) = (217°, 32°)$; the probability of finding that value or greater by chance is $7.9 \times 10^{-4}$.

The $a\cos\gamma_i$ fit for the actual handedness assignments gave $a = -0.0408$ with an uncertainty $\sigma_a = \pm 0.011$. Samples with randomly generated handedness and $(\alpha_A, \delta_A) = (217°, 32°)$ showed that $a/\sigma_a$ had a Gaussian distribution with an rms width of $a/\sigma_a = 1.001$.

The minimum in Fig. 3 is quite broad because the axis of the spirals could not be determined, only their handedness as ±1. The uncertainty in space angle is estimated to be ~35° as determined from the angle at which the $\chi^2$ probability doubles from its minimum value.

*Overall Asymmetry*–In the earlier analysis [5], in order to determine the overall statistical significance of the apparent asymmetry, I used data in the declination range $-19° < \delta < 60°$ and the right ascension ranges $195° \pm 45°$ and $0° \pm 40°$. Here I use the same $\delta$ range and right ascensions in a narrower range $195° \pm 30°$ and $15° \pm 30°$. Table I shows the resulting asymmetries and their uncertainties. The $165° < \alpha < 225°$ sector with 86% of the galaxies shows an asymmetry of $-0.0695 \pm 0.0127$, a $5.48\sigma$ effect. The data in the sparsely covered southern Galactic hemisphere ($-15° < \alpha < 45°$) show a small positive asymmetry as would be expected for a real signal. Overall the asymmetry is $-0.0607 \pm 0.0118$, a $5.15\sigma$ effect with a probability of $2.5 \times 10^{-7}$ for occurring by chance. The completely new data alone with $0.04 < z < 0.085$ gives an asymmetry $-0.0569 \pm 0.0142$ with a probability of $6.0 \times 10^{-5}$. For $z < 0.04$ the asymmetry is $-0.0692 \pm 0.0212$ with a probability of $1.1 \times 10^{-3}$. Thus there is no indication of a $z$ dependence of the asymmetry.

Table I. Number counts and net asymmetries $\langle A \rangle = (N_R - N_L)/N_{Tot}$ for the right ascension ranges indicated. The last two columns give the number of standard deviations for the $\langle A \rangle$ and the probability.

| $\alpha$ Range | $N_R$ | $N_L$ | $N_{Tot}$ | $\langle A \rangle \pm \sigma$ | $\langle A \rangle / \sigma$ | Prob. |
|---|---|---|---|---|---|---|
| -15° to 45° | 495 | 490 | 985 | 0.005±0.032 | +0.16 | 0.87 |
| 165° to 225° | 2890 | 3322 | 6212 | −0.0695±0.0127 | −5.48 | $2.1 \times 10^{-8}$ |
| Overall | | | 7197 | -0.0607±0.0118 | −5.15 | $2.5 \times 10^{-7}$ |

## 4. Comparison with other studies

*Iye and Sugai* – Iye and Sugai [10] have published a catalog of spin orientations of galaxies in the southern Galactic hemisphere that contains 8287 spiral galaxies. Of these, 3118 had *R* or *L* handedness about which both scanners agreed. I have analyzed their catalog using the sector $-15° < \alpha < +45°$ and $-60° < \delta < +5°$, directly opposite that used above[2]. Redshifts of most of their galaxies were not measured, so only their $(\alpha, \delta)$ were used. This gave an asymmetry $+0.047 \pm 0.029$ with a preponderance of right-handed spirals in the southern Galactic hemisphere, in ex-cellent agreement with the asymmetry $|A| = 0.0695\pm0.0127$ that I observe for $165°<\alpha<225°$ with a preponderance of left-handed spirals. This provides an independent confirmation of a spin asymmetry at the 1.6$\sigma$ level. Without ($\alpha, \delta$) cuts their overall asymmetry was $0.000\pm0.014$, consistent with no bias in their study. The combined probability for the dipole term found here and the asymmetry found for the Iye and Sugai catalog is $0.205(7.9\times10^{-4}) \sim 1.6\times10^{-4}$.

---

[2] Note, however, that the Iye-Sugai catalog does not contain galaxies with declinations $\delta > -18°$.

*Galaxy Zoo* – Galaxy Zoo (Lintott et al. [11]) is an online project in which >100,000 volunteers visually classify the morphologies of galaxies selected from the spectroscopic sample of the SDSS DR6, the same sample used here. In Land et al. [12] they investigate the possibility of a large scale spin anisotropy. Each galaxy was classified an average of 39 times. Those galaxies for which over 80% of the votes agreed constituted their "clean" sample and over 95% their "superclean" sample. They found that there was a large $L/R$ bias in their samples. Their clean sample contained 18471 $L$ (clockwise) spirals and 17100 $R$; the superclean sample contained 7034 $L$ and 6106 $R$ spirals[3]. This gives an asymmetry (bias?) of $-0.0385\pm0.0053$ for the clean sample and a much larger bias of $-0.0706\pm0.0087$ for the superclean sample that presumably contained more clearly recognizable spirals. This should be compared to the upper limit of $-0.0077 \pm 0.0062$ for the bias found in this study as discussed in the section on biases above. Land et al. attributed these biases to the design of the Galaxy Zoo website or to a human pattern recognition effect that was shared by all their volunteers. A later bias study, mainly with the superclean sample, that compared monochrome images with mirrored RGB images found similar biases. They corrected for the bias by requiring only 78% agreement between scanners for the $R$ galaxies in the clean sample and 94% in the superclean. They assumed the bias was independent of redshift and magnitude, despite the fact that it is much easier to correctly assign the handedness of larger, brighter galaxies. In their analysis, they found a dipole term of about 2$\sigma$ along ($\alpha, \delta$)=(161°,11°) consistent with the axis I found in [5] at (202°,25°). When a monopole (bias) term was also allowed, this became a 1$\sigma$ effect.

It is difficult to compare this study with the Galaxy Zoo one. They used galaxies with redshifts up to 0.3, whereas I used those with $z<0.085$ and restricted the magnitude range because of the difficulty in assigning the handedness of fainter and less well resolved galaxies. Their large biases also caused large uncertainties in the monopole/bias term, while the biases in this analysis are consistent with 0.

---

[3] See also Table 2 of Lintott et al. [11]. The uncertainties in the asymmetry are calculated as $\sigma(A) = 1/\sqrt{R+L}$.

## 5. Asymmetry vs. galaxy separation

It is instructive to study how the asymmetry varies with galaxy separation in order to see if the apparent *L-R* asymmetry is truly a long-range correlation, rather than a correlation with a length scale on the order of galaxy cluster sizes.

Figure 5(a) shows the distribution of spiral pair separations in the sample, assuming $\Omega_m=0.3$. This was generated by binning the separation of all pairs of spirals and taking the ratio to that of the separation of spirals with their $\alpha$'s, $\delta$'s, and redshifts scrambled among the sample of spirals. The scrambling effectively destroys the spatial correlations of the sample for small separations. The ratio (NOT_SCRAMBLED/SCRAMBLED) − 1 is plotted. The effect of clustering is apparent for separations less than ~10 Mpc/h.

Figure 5(b) shows the asymmetries (*LL*–*RR*)/(*LL*+*RR*) as + signs and that of (*LL*–0.5\**LR*)/ (*LL*+0.5\**LR*) as o's, where *LL* is the number of Left-Left pairs in a particular separation bin; *RR* is the number of Right-Right; and *LR* is the number of Left-Right or Right-Left. (The *LR* pairs have twice the probability of the *LL* and *RR*, as in coin flips. This is the reason for the 0.5 in front of the *LR* term.) The preponderance of *LL* pairs is clear out to separations ~210 Mpc/h. As might be expected for an unbiased sample, the asymmetry between *LL* and *LR* pairs is half that for *LL* and *RR* pairs. For separations less than ~20 Mpc the correlations decrease significantly. This shows that the *LL* excess is truly a long-range phenomenon, extending over almost the whole sample. The decrease at small separations is due to the fact that the spins of nearby spirals are likely to be opposite because near collisions of spiral galaxies tend to spin them in opposite directions due to angular momentum conservation. In other words, nearby spiral galaxies are more likely to have their spins antiparallel than more distant pairs.

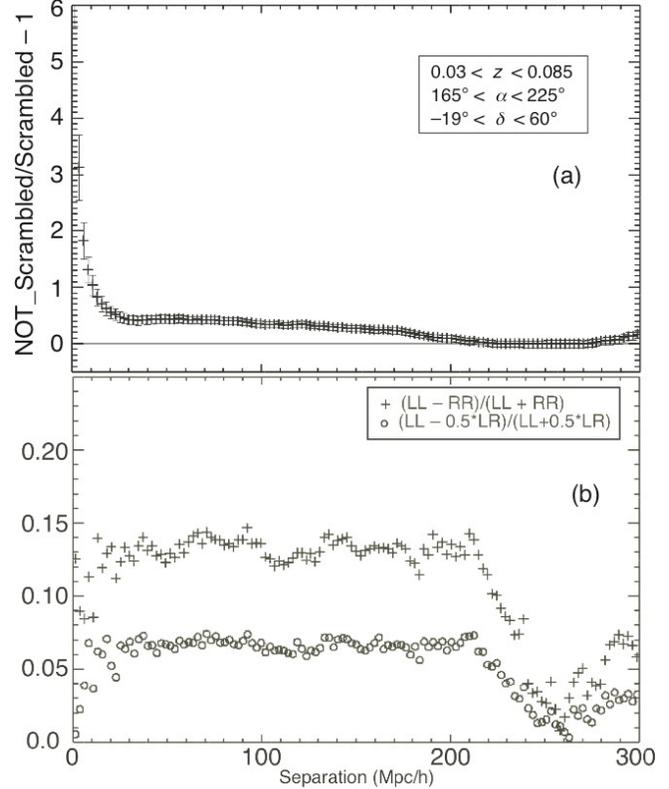

FIG. 5(a)–Spatial correlation of the spiral galaxies in the sample vs. 3D separation. The peaking for separations <20 Mpc/h is due to clustering. (b) The asymmetry of spiral galaxy spins vs. separation. The correlation extends out to separations ~210 Mpc/h.

## 6. Discussion

This analysis used data selection cuts very similar to those in the previous analysis [5], and the sample contains about 7 times as many galaxies, mainly due to its extension from $z=0.04$ to $z=0.085$. As discussed in Sect. 3, the unbinned analysis with no a priori assumptions of a dipole axis shows the probability of finding the observed dipole term by chance is $7.9 \times 10^{-4}$. If the same asymmetry axis found in [5] is assumed, the new analysis for $z<0.04$ gives an asymmetry $-0.0692\pm0.0212$ with a probability of $1.1 \times 10^{-3}$, and the completely new, statistically independent, data with $0.04 < z < 0.085$ gives an asymmetry $-0.0569\pm 0.0142$ with a probability of $6.0 \times 10^{-5}$. The combined probability for these two statistically independent samples is $6.6 \times 10^{-8}$. The Iye-Sugai spin orientation catalog in the southern Galactic hemisphere provides independent evidence of a spin asymmetry at the $1.6\sigma$ level.

Thus, the statistical evidence for a spin asymmetry is very strong. In addition, there are strong arguments against the possibility that the observed asymmetry is due to systematic effects. It is hard to imagine any systematic effect in the SDSS survey itself that would preferentially change right-handed spirals into left-handed. The only plausible systematic would be a scanner bias to preferentially classify spirals as left-handed. In this analysis, this possibility was effectively eliminated by randomly mirroring half of the galaxies as presented to the scanners with no visual cues as to the mirroring. The overall sample was consistent with no asymmetry at the 1% level. In the original analyses [5], all of the scanning was done by the author or a rotating mask algorithm. In this study >95% of the scanning was done by the undergraduate scanners, yet the results were in very good agreement. The fact that the sparse SDSS data in the southern Galactic hemisphere and the Iye-Sugai catalog gave asymmetries consistent with the opposite sign is also some evidence against a bias toward left-handed spirals.

The axis of the asymmetry lies near $(\alpha_A, \delta_A) = (217°, 32°)$, or $(l, b) = (52°, 68.5°)$ in Galactic coordinates with the sense of the axis defined to be along the direction of the $L$ (↺) excess. The uncertainty in space angle of the axis is ~35°. This axis is 21.5° away from the north pole of our galaxy (NGP in Fig. 2) at $b=90°$. Thus, our galaxy has its spin vector generally aligned with the preferred axis of spiral galaxies, corresponding to a probability for this to occur by chance of $(1-\cos 21.5°)/2 = 3.5\%$. There is no obvious redshift dependence of the asymmetry out to $z \approx 0.085$, well beyond the scale of superclustering[4]. The spin correlation extends out to separations ~210 Mpc/h, while galaxies with separations < 20 Mpc/h have smaller spin correlations.

This constitutes strong evidence for a violation of the Cosmological Principle, as well as a parity violation on cosmic scales.  Extension of this study to larger redshifts will be difficult due to the problem of reliably recognizing the handedness of faint galaxies.  A more complete survey and analysis of spiral galaxies in the southern galactic hemisphere, comparable to the SDSS, would go a long way toward verifying the asymmetry.

---

[4] For example, Virgo, at the center of our local supercluster, is at a redshift ~0.004.

There is now a vast literature on possible observations of cosmic anisotropies that suggest a preferred axis.  An excellent recent summary is given by L. Perivolaropoulos [13].  Many of these revolve around apparent anomalies in the WMAP one-year, three-year, and five-year data.  For example, analyzing the one-year data, Land and Magueijo [14] find an unlikely alignment of the low $l$ multipoles and a correlation of azimuthal phases between $l = 3$ and $l = 5$ with an apparent axis $(l, b) = (260°, 60°)$, an alignment they refer to as the "axis of evil".  Using the three-year WMAP data, Copi et al. [15] find a similar correlation of low $l$ multipoles and a significant lack of correlations for scales >60°.  Analyzing the WMAP five-year and three-year data, Bernui [16] finds a significant asymmetry in large-angle correlations between the north and south Galactic hemispheres at >90% confidence level, depending on the map used.  Su and Chu [17], in a recent reanalysis of the WMAP data, find a general alignment of the directions for $l = 2$ to 10 modes to within about $1/4^{th}$ of the northern Galactic hemisphere at latitudes between about 45° and 85°.  Similarly, Råth et al. [18] find evidence for asymmetries and non-Gaussianities between the North and South Galactic polar regions.  In a recent analysis of the WMAP dipole asymmetry, Hoftuft et al. [19] find nonzero dipole amplitudes $>3\sigma$ in each of several wavelength bands with an axis at $(l, b) = (44°, 22°) \pm 24°$,  Samal et al. [20] rule out isotropy in the combined CMB to confidence levels of better than 99.9%.  There is also a considerable literature on other anisotropic effects in polarizations of radio galaxies and the optical frequency polarization correlation of quasars.  References [2, 3] discuss observations of apparent parity-violating effects.

Perivolaropoulos[13] summarizes six independent pieces of evidence for a preferred axis with a mean axis at $(l, b) = (278° \pm 26°, 45° \pm 27°)$ or $(\alpha, \delta) = (175°, -14°)$.  Though this is about 60° away from the spiral axis, it is outside of the SDSS coverage.  Therefore, the present study does not have much sensitivity to a preferred spiral axis in this region. [5]

I am indebted to Dr. M. Iye for providing the spin catalog of southern galaxies.  Undergraduates E. Mallen, A. Bomers, J. Middleton, and M. Pearce made important contributions in their diligent work as scanners.  B. McCorkle did the JAVA/HTML programming.  This work would not have been possible without the dedicated efforts of the SDSS collaboration.  An anonymous reviewer made many helpful suggestions.

---

[5] A file with the spin assignments, coordinates, and magnitudes is available as supplementary material.